\documentclass[english,aip]{revtex4}

\makeatletter 
\usepackage{amssymb} 
\usepackage[colorlinks=true,linkcolor=blue]{hyperref}
\usepackage{amsmath}
\usepackage{graphicx}
\usepackage{color}
\usepackage{ulem}

\begin{document}

	\title{RT-TDDFT study of hole oscillations in B-DNA monomers and dimers}

	\author{M. Tassi}
	\email[Corresponding author: ]{mtassi@phys.uoa.gr}
	\author{A. Morphis} 
	\author{K. Lambropoulos}
	\author{C. Simserides}
	\affiliation{National and Kapodistrian University of Athens, Department of Physics, Panepistimiopolis, Zografos, GR-15784, Athens, Greece}

	\begin{abstract}
		We employ Real-Time Time-Dependent Density Functional Theory to study hole  oscillations within a B-DNA monomer (one base pair) or dimer (two base pairs). 
		Placing the hole initially at any of the bases which make up a base pair, results in THz  oscillations, albeit of negligible amplitude. 
		Placing the hole initially at any of the base pairs which make up a dimer is more interesting:
		For dimers made of identical monomers, we predict oscillations with frequencies in the range $f \approx$ 20-80 THz, with a maximum transfer percentage close to 1. 
		For dimers made of different monomers, $f \approx$ 80-400 THz, but with very small or small maximum transfer percentage. 
		We compare our results with those obtained recently via our Tight-Binding approaches and find that they are in good agreement.
	\end{abstract}
	
	\maketitle

	


\section{Introduction} 
\label{sec:introduction} 
Carriers can be inserted in DNA via electrodes or generated by UV irradiation or by oxidation and reduction. We had better discriminate between the words \textit{transport} (implying the use of bias between electrodes), \textit{transfer} (a carrier moves without bias) and \textit{migration} (a transfer over rather long distances). Charge transport, transfer and migration   in DNA are affected by many external factors such as chemical environment, interaction with the substrate or quality of contacts~\cite{Macia:2009}, as well as by many intrinsic factors like the sequence of bases~\cite{Cuniberti:2007}.
Charge transfer in DNA is important for DNA damage and repair~\cite{Dandliker:1997,Rajski:2000,Giese:2002,Giese:2006} and it may be used as an indicator for the separation between cancer tumor and healthy tissue~\cite{Shih:2011}. Although (unbiased) charge transfer in DNA nearly vanishes after 10 to 20 nm \cite{Simserides:2014,LChMKTS:2015,LChMKLThTS:2016}, DNA is still a promising component for (biased) charge transport in molecular electronics, e.g., as a short molecular wire \cite{Wohlgamuth:2013}. It may also be exploited in nanotechnology for nanosensors \cite{Huang:2016} and molecular wires~\cite{GenereuxBarton:2010,KawaiMajima:2015}.

The scope of this article is to study the effect of base sequence on charge transfer, in all possible monomers (i.e. within one base pair) and dimers (i.e. within two base pairs), using Real-Time Time-Dependent Density Functional Theory (RT-TDDFT) \cite{Theilhaber:1992,YabanaBertsch:1996}. In previous articles, we studied small and long B-DNA segments~\cite{Simserides:2014,LKGS:2014,LChMKTS:2015,LKMTLGThChS:2016,LChMKLThTS:2016} with Tight Binding (TB) approaches, specifically with a wire model and an extended ladder model. 
We determined the spatio-temporal evolution of an extra carrier (hole or electron) along a $N$ base-pair DNA segment and showed that the carrier movement has frequency content mainly in the THz domain, i.e., depending on the case, in the NIR, MIR and FIR range \cite{ISO}.
This part of the electromagnetic (EM) spectrum is significant because it can be used to extract complementary to traditional spectroscopic information e.g. about hydrogen bond stretching, bond torsion in liquids and gases, low-frequency bond vibrations and so on, and because it is relatively non-invasive compared to higher-frequency regions of the EM spectrum~\cite{EM}.

In this article we study B-DNA monomers and dimers with RT-TDDFT,  which is one of the few computationally viable techniques to model carrier dynamics in molecular systems \cite{Marques:2006,Liang:2011,LopataGovind:2011,Ullrich:2012}.
RT-TDDFT can simulate density fluctuations of the order of attoseconds to femtoseconds and can model the effects of an ultrafast, intense, short-pulse laser field, including the full (nonlinear) response of the electron density \cite{ProvorseIsborn:2016}. RT-TDDFT also provides the full frequency dependence of properties of interest via Fourier transform from the time domain to the frequency domain. In addition, the determination of many-particle eigenstates is avoided, and the calculation is reduced to an independent particle problem carried out in real time \cite{Takimoto:2007}. Recently, RT-TDDFT was applied \cite{Partovi-Azar:2017} to study the charge transfer between the oxidized or reduced a-sulfur monomers  and the neighbouring neutral ones. Another recent work used RT-TDDFT to study electron transfer through oligo-p-phenylenevinylene (OPV) and carbon bridged OPV (COPV) \cite{Granasetal:2017}. 

The rest of this article is organized as follows. In Section \ref{sec:theory} we sketch the basic formalism of RT-TDDFT. In Section \ref{sec:details} we give some computational details and explain our notation. In Section \ref{sec:results} we present our results. Finally, in Section \ref{sec:conclusion} we state our conclusions.

\section{RT-TDDFT Formalism} 
\label{sec:theory} 
Density Functional Theory (DFT) \cite{HohenbergKohn:1964,KohnSham:1965} is an efficient method for treating ground state properties of many electron systems e.g. molecules or solids. Years after its development, it was extended \cite{RungeGross:1984} to time dependent systems (TDDFT). Specifically, the Time-Dependent Kohn-Sham (TDKS) equations with an effective potential energy $\upsilon_\textrm{KS}(\mathbf{r},t)$, uniquely described by the time-dependent charge density, $\rho(\mathbf{r},t)$, are, in atomic units,
\begin{eqnarray}  \label{TDKS}
 i \frac{\partial}{\partial t}\Psi_{j}(\mathbf{r},t) =
\bigg[ -\frac{1}{2} \nabla^{2} + \upsilon_\textrm{KS}(\textbf{r},t) \bigg] \Psi_{j}(\mathbf{r},t) = \\ \nonumber
\bigg[ -\frac{1}{2} \nabla^{2} + \upsilon_\textrm{ext}(\textbf{r},t)
                               + \upsilon_\textrm{H}  (\textbf{r},t)
                               + \upsilon_\textrm{xc}[\rho](\textbf{r},t) \bigg] \Psi_{j}(\mathbf{r},t).
\end{eqnarray}
The charge density is the sum over all occupied orbitals $j = 1, 2, \dots N_{occ}$, i.e.,
\begin{equation}
\rho(\textbf{r},t) = \sum_{j=1}^{N_{occ}} |\Psi_{j}(\mathbf{r},t)|^2.
\end{equation}
$\upsilon_\textrm{ext}(\textbf{r},t)$ includes external fields and nuclear potentials,
$\upsilon_\textrm{H}(\textbf{r},t)$ is the Hartree potential energy.
Exchange and correlation effects are included in
$\upsilon_\textrm{xc}[\rho](\textbf{r},t)$.

RT-TDDFT is based on a direct numerical integration of Eq.~\ref{TDKS}.
This differs from the traditional linear-response approach,
which is not actually a time-resolved method but instead solves Eq.~\ref{TDKS}
in the frequency domain for the excitation energies of a system subject to a small perturbation \cite{LopataGovind:2011}.
Within RT-TDDFT, we solve the TDKS equations and obtain the electron density at each time step. The electron density is then used for the calculation of the Hamiltonian in the next cycle of the self-consistent process.

\section{Computational details and notations} 
\label{sec:details} 
With the notation YX we mean that the bases Y and X of two successive base pairs Y-Y$_\textrm{compl}$ and X-X$_\textrm{compl}$ are located at the same strand in the direction $5'$-$3'$.
X$_\textrm{compl}$ (Y$_\textrm{compl}$) is the complementary base of X (Y).
In other words, the notation YX means that Y-Y$_\textrm{compl}$ is the one base pair and X-X$_\textrm{compl}$ is the other base pair, separated by 3.4 {\AA} and twisted clockwise by 36$^\textrm{o}$ relatively
to the first base pair, along the growth axis of the nucleotide chain.
For example, the notation AC denotes that one strand contains A and C in the direction $5'$-$3'$ and the complementary strand contains T and G in the direction $3'$-$5'$. The notation (10) means that the hole is initially placed at the 1st base pair of the dimer and the notation (01) means that it is initially placed at the 2nd base pair of the dimer.

For our DFT and RT-TDDFT calculations we used the NWChem open-source computational package \cite{nwchem:2010}.
Additionally, we performed Fourier analysis, implemented with MATLAB, to determine the frequency content of the charge oscillations.

The geometry of monomers and all dimers  was produced via the BIOVIA Discovery Studio \cite{biovia}. The backbone was removed and H atoms were added at the corresponding atomic sites.

The range-separated functional CAM-B3LYP~\cite{Yanai:2004}, which is appropriate for the correct estimation of the exchange energy, both at short and long ranges, was used for all monomers and dimers. The calculations were performed using the 6-31++G** \cite{Hehre:1972}, 3-21++G \cite{Binkley:1980} and aug-cc-pVDZ \cite{Dunning:1989} basis sets, which  include diffuse functions, for all systems. 
Within a small number of time steps, we noticed that our results were similar (cf. Appendix). Hence, we chose to perform our longer simulations with the 6-31++G** basis set \cite{Hehre:1972}, that is, the least computationally expensive between the converging sets.


In a Gaussian basis set, it is most natural to use the single particle reduced density matrix, whose time evolution is governed by the von Neumann equation. The Magnus propagator is used in NWChem's RT-TDDFT implementation, which is both stable and conserves the density matrix idempotency \cite{LopataGovind:2011}. At the end of each time step the fragments' charge is calculated with an appropriate population analysis method, along with the total dipole moment.

For monomers, the initial state (i.e. hole localized at a specific base) was produced by CDFT, where the charge constraint was calculated with L\"{o}wdin population analysis that is also used in the subsequent RT-TDDFT simulation. For dimers, the initial state (i.e. charge localization) was produced by the following procedure: first, a ground state DFT calculation of the charged and neutral isolated monomers was performed, yielding the corresponding monomers' eigenstates. Then, the dimer's eigenstates were approximated in a non self-consistent manner (NOSCF) by the orthogonalized monomers' eigenstates.
Since the monomers comprising each dimer are 3.4 {\AA} apart, this is a fairly good approximation that also helps circumvent CDFT convergence problems.

L\"{o}wdin \cite{Loewdin:1950} population analysis was integrated into RT-TDDFT module of NWChem for the calculation of each monomer's charge at each time step. L\"{o}wdin population analysis is known to be much less basis-set dependent and also it does not suffer from the ultra-fast charge oscillations that Mulliken analysis (which is the default scheme in NWChem's RT-TDDFT) artificially introduces in RT-TDDFT charge simulations. 
As a result, L\"{o}wdin population analysis gives a more clear picture of charge oscillations convergence towards the basis-set limit, cf. Figs.~\ref{fig:same}-\ref{fig:diff}.
The main frequencies of charge oscillations are extracted from the results via Fourier analysis. In order to increase the resolution, zero-padding with appropriate signal attenuation was used where necessary.

\section{Results} 
\label{sec:results} 
Before discussing our RT-TDDFT results, we give a brief summary of the TB picture. In TB, charge transfer between the two bases of a monomer or between the two base pairs of a dimer can be easily analytically solved \cite{Simserides:2014, LKGS:2014, LKMTLGThChS:2016}.
In both cases, it is a movement between the two sites $i,j$ of a two-site system and it is  mathematically similar to Rabi oscillations in Quantum Optics \cite{QOL:2016}. 
The maximum transfer percentage $p$, i.e., the maximum probability to find the carrier at the site where it was not initially placed is 
\begin{equation}\label{p}
p = \frac{(2t_{ij})^2}{(2t_{ij})^2+(\Delta_{ij})^2}, 
\end{equation}
where $\Delta_{ij}$ is the difference between the two sites' on-site energies and $t_{ij}$ is the hopping parameter between the two sites. Since between the two bases of a monomer, $\Delta_{ij}$ is much larger than $t_{ij}$, $p$ is negligible. For movements between the two monomers of a dimer things are different: For dimers made of identical monomers, since $\Delta_{ij}=0$, it follows that $p = 1$. For dimers made of different monomers $p < 1$; in fact, it is usually negligible with one exception, a hole moving between the two monomers of a GA dimer, where $p$ is of the order of 0.25. In TB, the frequency and period of charge oscillations is given by 
\begin{equation}\label{fandT}
f = \frac{1}{T} = \frac{\sqrt{(2t_{ij})^2 + \Delta_{ij}^2}}{h},
\end{equation}
hence, increasing $t_{ij}$ and / or  $\Delta_{ij}$ results in higher frequencies.
In what follows, this simple picture is qualitatively obeyed by our RT-TDDFT results.

Hole oscillations between the two bases of a base pair are in the THz regime, but have negligible maximum transfer percentage, in agreement with the TB results \cite{LKMTLGThChS:2016}. 

\begin{figure}[h!]
\centering
\includegraphics[width=10cm]{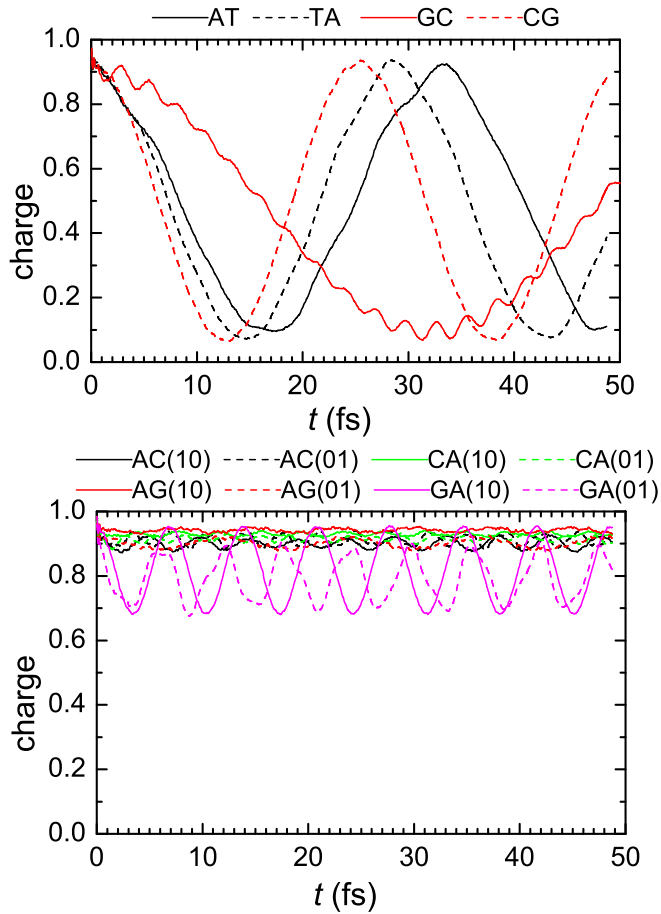}
\caption{Net charge versus time for the monomer where the hole was initially placed: (upper panel) dimers made of identical monomers, (lower panel) dimers made of different monomers. The notation (10) means that the hole is initially placed at the 1st base pair of the dimer, while the notation (01) means that the hole is initially placed at the 2nd base pair of the dimer.}
\label{fig:osci}
\end{figure}

Hole oscillations between the two base pairs of a dimer, with the 6-31++G** basis set, are shown in Fig \ref{fig:osci}. We place the hole initially at one of the monomers which make up the dimer and call $p$ the maximum transfer percentage to the other monomer.
For dimers made of identical monomers (AT, TA, GC, CG) the hole is transferred almost completely to the other monomer ($p \approx$ 0.9). For dimers made of different monomers (AC, CA, GA, AG), $p$ is close to zero with the exception of the GA dimer, where $p \approx$ 0.3. Our calculations for the cases AT(01), TA(01), GC(01), and CG(01) are not included in Fig.\ref{fig:osci}, since they are identical with the cases AT(10), TA(10), GC(10), CG(10), respectively, as expected from base-pair symmetry and found also in Refs.~\cite{Simserides:2014,LKGS:2014,LKMTLGThChS:2016}. 

\begin{figure}[h!]
\centering
\includegraphics[width=10cm]{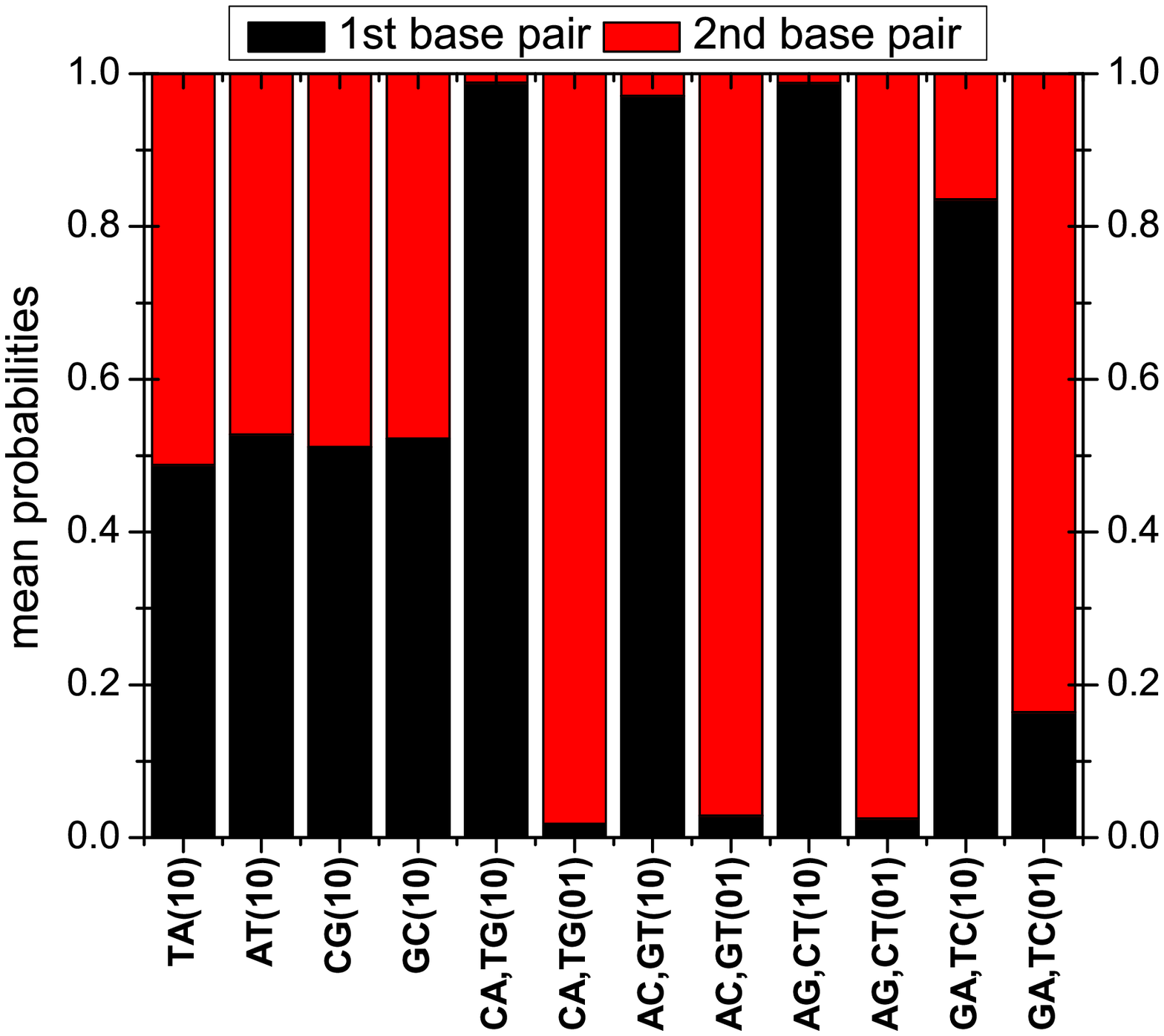}
\caption{The mean probabilities to find the extra hole, at each base pair of a dimer,  having placed it initially at the 1st (10) or at the 2nd (01) base pair.}
\label{fig:dimer_mean_homo}
\end{figure}

In Fig.~\ref{fig:dimer_mean_homo} we present the mean probabilities to find the extra hole  at each monomer, having placed it initially either at the 1st monomer (10) or at the 2nd monomer (01). Let us call $\mathcal{P}_i$ the mean probability at the monomer where the hole is initially placed, and $\mathcal{P}_f$ the mean probability at the other monomer. We observe that for dimers made of identical monomers, where $p \approx 1$, we have $\mathcal{P}_i \approx \mathcal{P}_f \approx 0.5$, while for dimers made of different monomers $\mathcal{P}_i\approx$ $1$ and $\mathcal{P}_f \approx 0$, with the exception of the GA dimer. This, again, agrees qualitatively with the TB picture \cite{Simserides:2014,LKGS:2014,LKMTLGThChS:2016}.

\begin{figure}[h!]
\centering
\includegraphics[width=10cm]{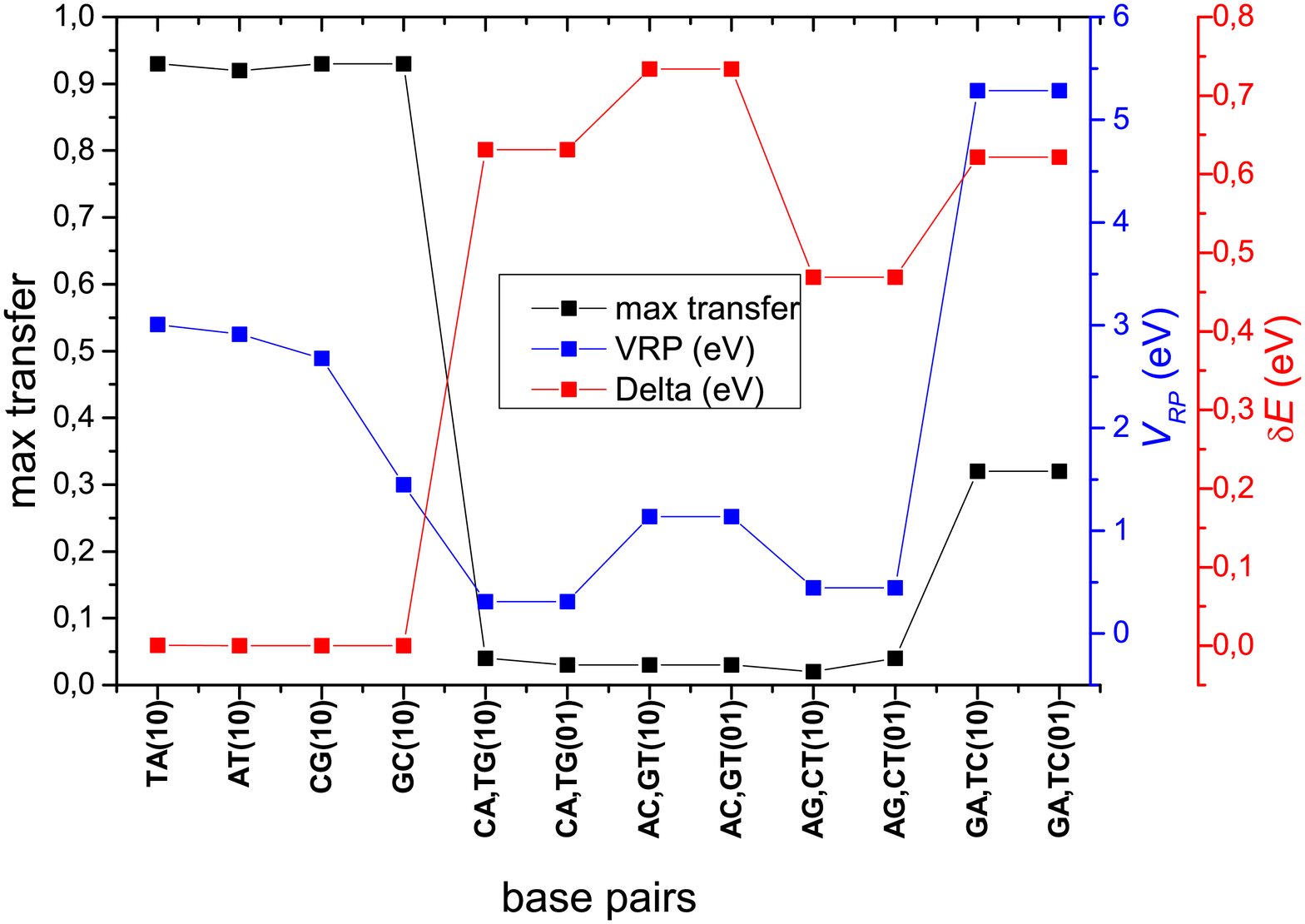}
\caption{Maximum transfer, i.e., oscillation amplitude $p$, electron transfer coupling energy $|V_{\textrm{RP}}|$ and energy difference $\delta E$ between reactant and product states. Initially, the hole is placed at the 1st monomer (10) or at the 2nd monomer (01).}
\label{fig:transfer_homo}
\end{figure}

In Fig.~\ref{fig:transfer_homo} we depict the maximum transfer percentage $p$  as well as the electronic coupling energy (also called the electron transfer matrix element) $V_{\textrm{RP}}$ between the reactant and product states $R$ and $P$ \cite{NewtonSutin:1984,MarcusSutin:1985,Bolton:1991,Farazdel:1990}, which reflects the magnitude of the interaction between the two monomers. 
In our case reactant state corresponds to the hole at the initial placement and product state corresponds to the hole at the other monomer. 
We also show the energy difference $\delta E$ between reactant and product states. 
In analogy with a Rabi oscillation, we expect charge transfer to increase increasing $V_{\textrm{RP}}$ and to decrease decreasing $\delta E$. Additionally, for  $\delta E \approx 0$, we expect $p$ to be close to 1.
For the AT, TA, GC and CG dimers, $\delta E \approx 0$, hence $p$ is indeed close to $1$. For dimers made of different monomers, we observe that AC, CA and AG have large $\delta E \approx$ 0.45-0.75 eV and small $V_{\textrm{RP}} \approx$ 0-1 eV, therefore their $p$ is insignificant. An exception is the dimer GA which not only has a large $\delta E \approx$ 0.65 eV, but also the largest $V_{\textrm{RP}}$ of all dimers $\approx$ 5 eV which makes charge transfer significant, with $p$ $\approx 0.3$.

\begin{figure}[h!]
\centering
\includegraphics[width=10cm]{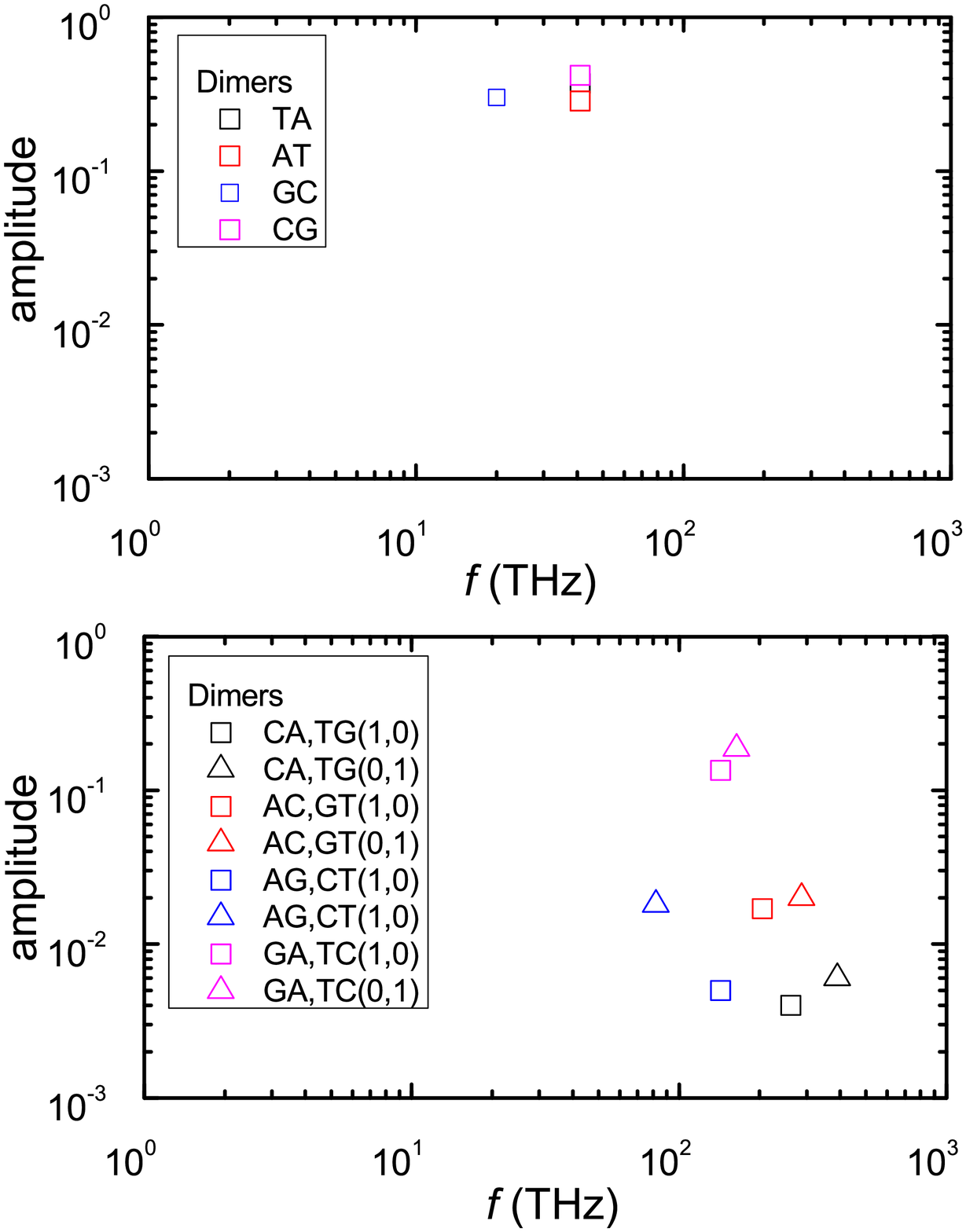}
\caption{Frequencies of hole oscillations in dimers made of identical monomers (upper panel) and different monomers (lower panel), obtained via Fourier analysis.}
\label{fig:fourier_dimers}
\end{figure}

\begin{figure}[h!]
\centering
\includegraphics[width=10cm]{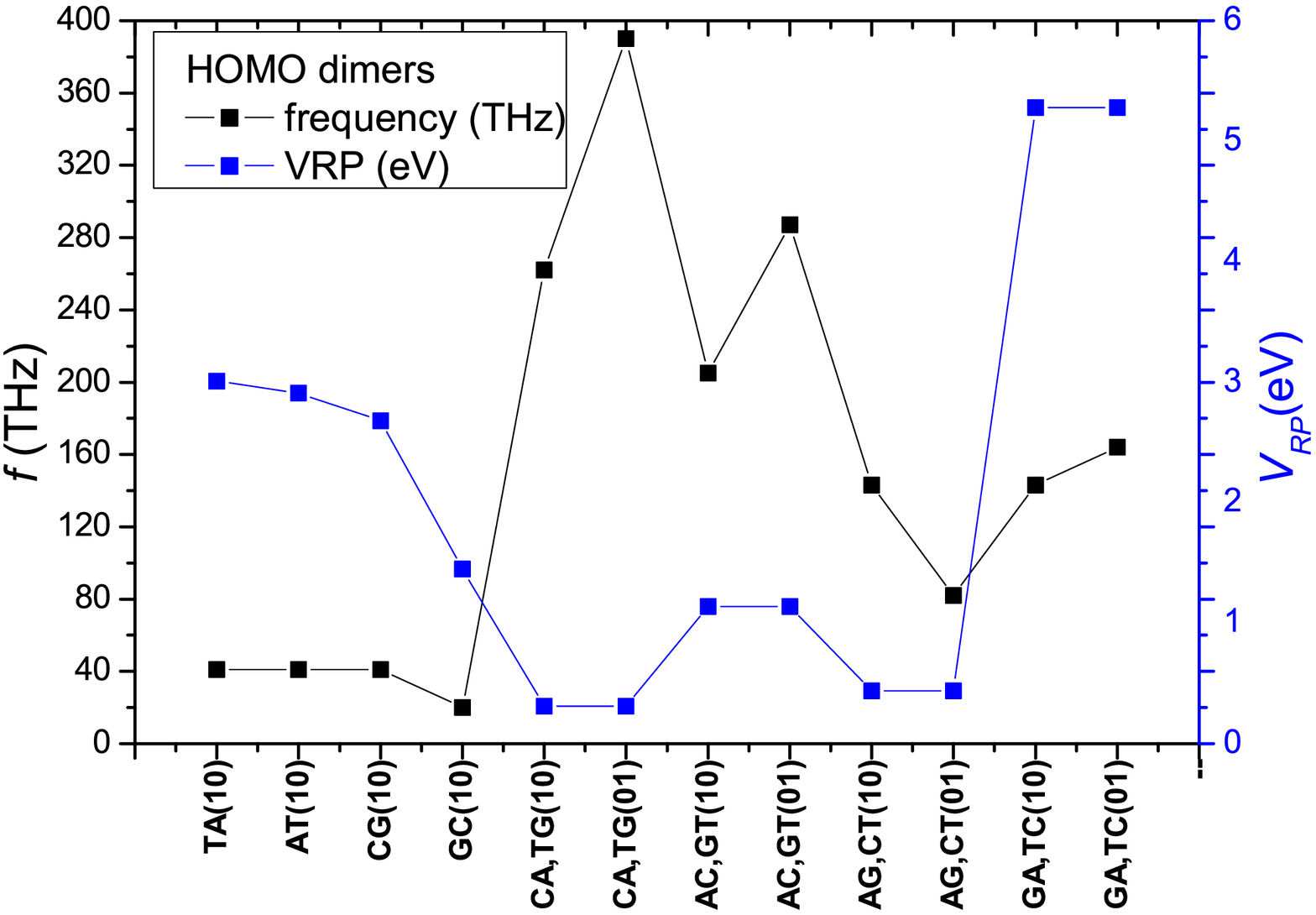}
\caption{Extra hole oscillation frequency versus $|V_{\textrm{RP}}|$ for all dimers.}
\label{fig:fvsvrp}
\end{figure}

In Fig. \ref{fig:fourier_dimers} we present the frequencies of hole oscillations obtained by Fourier analysis. For dimers made of identical monomers (upper panel) the frequencies $f \approx$ 20-80 THz and their amplitudes $\approx$ 0.2-0.5. For dimers made of different monomers (lower panel) $f \approx$ 80-400 THz and the amplitudes $\approx$ 0.003-0.02, except for the two GA cases which have amplitudes $\approx$ 0.1-0.2. Obviously, larger amplitudes reflect larger transfer. As expected by TB, for dimers made of different monomers we generally get higher frequencies than for dimers made of identical monomers.
Additionally, in Fig.~\ref{fig:fvsvrp} we observe that for dimers made of identical monomers the frequencies follow $|V_{\textrm{RP}}|$, which does not hold for dimers made of different monomers. This, again, agrees with the TB prediction (Eq.~\ref{fandT}). 

\begin{figure}[h!]
\centering
\includegraphics[width=8cm]{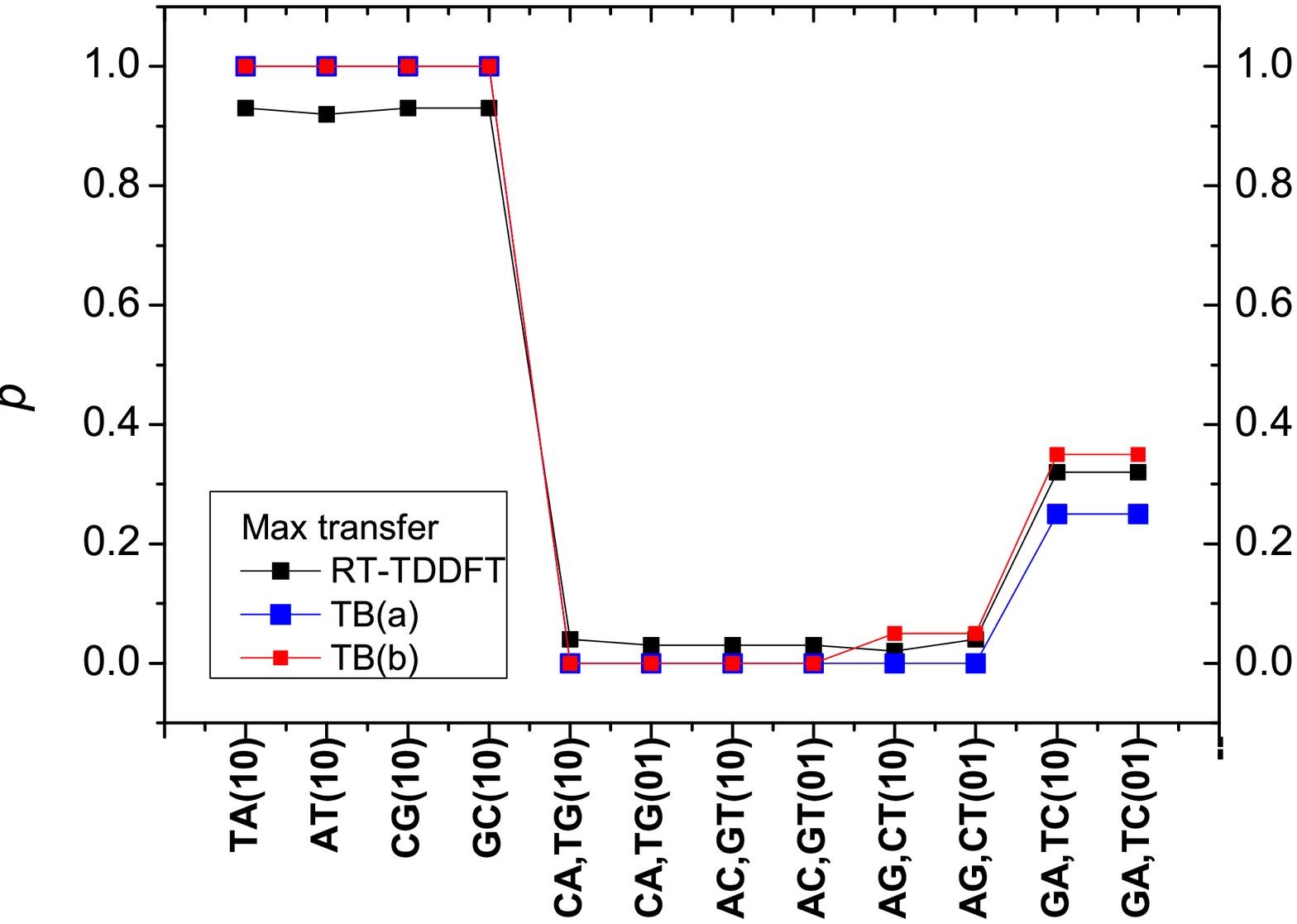}
\includegraphics[width=8cm]{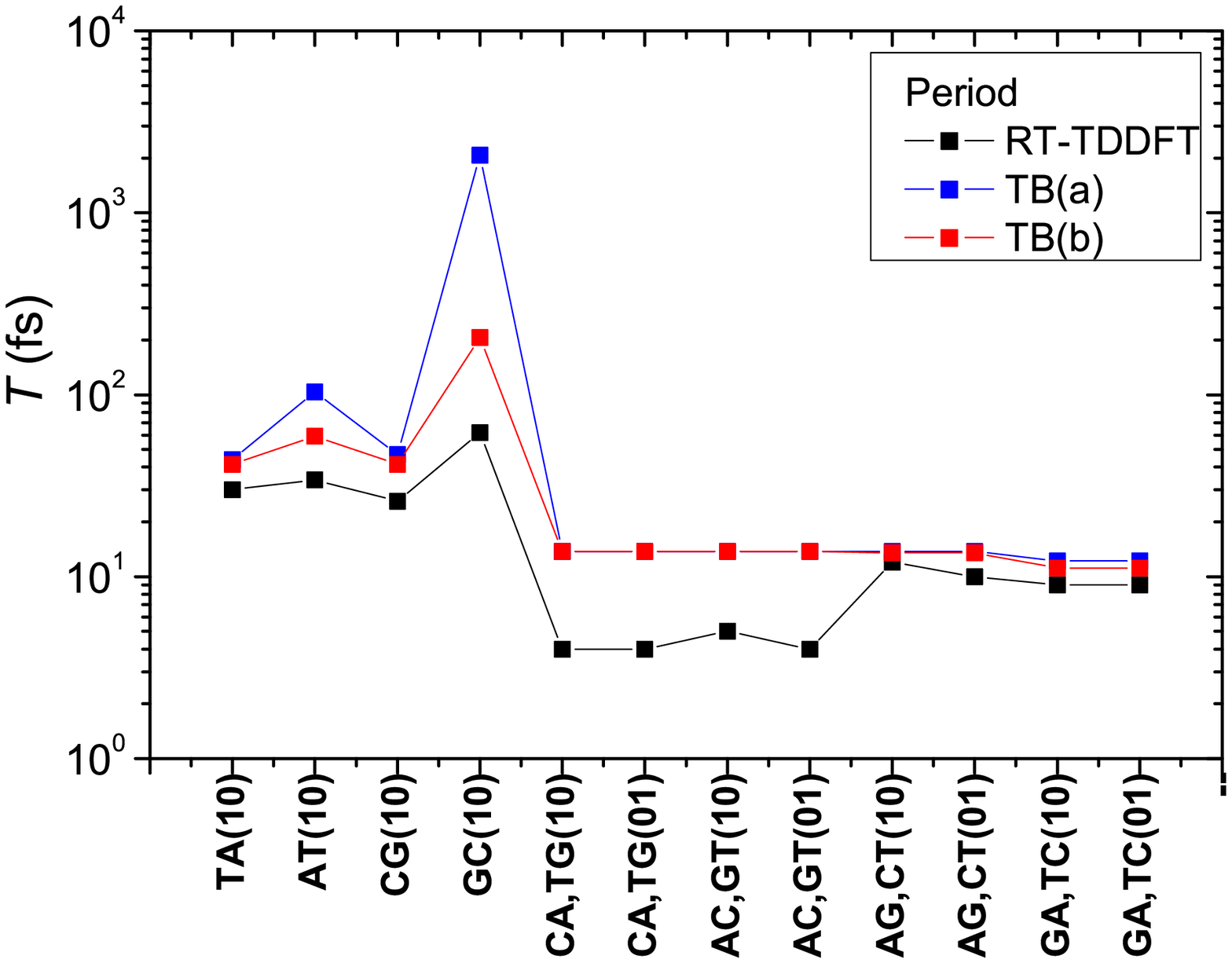}
\caption{Comparison of RT-TDDFT (this work) and TB from (a) Ref.~\cite{LKMTLGThChS:2016}  and (b) Ref.~\cite{Simserides:2014}: maximum transfer percentage (left panel) and oscillation periods (right panel).}
\label{fig:rtvstb}
\end{figure}

Finally, in Fig.~\ref{fig:rtvstb} we compare our RT-TDDFT results with those of TB \cite{Simserides:2014,LKMTLGThChS:2016} for two different sets of TB parameters: (a) used in Ref.~\cite{LKMTLGThChS:2016} (calculated in Ref.~\cite{HKS:2010-2011}) and (b) used in Ref.~\cite{Simserides:2014}. We observe that the maximum transfer percentages obtained by RT-TDDFT are in good agreement with those obtained by TB. For the periods, the results, both for RT-TDDFT and TB are quite close especially when we compare them with parametrization (b) used in Ref.~\cite{Simserides:2014}.
 \clearpage

\section{Conclusion} 
\label{sec:conclusion} 
We studied how base sequence affects hole transfer in small DNA segments (between the two bases of a monomer and between the two base pairs of a dimer) with the RT-TDDFT method.  

We calculated the maximum transfer percentage, i.e., the oscillation amplitude, $p$, the mean probability at the site where the hole is initially placed, $\mathcal{P}_i$, and the mean probability at the other site, $\mathcal{P}_f$, as well as the frequency content of the oscillations.

Hole oscillations between the two bases of a base pair are in the THz regime, but have negligible maximum transfer percentage.

For dimers made of identical monomers, the maximum transfer percentage is almost $1$. 
For dimers made of different monomers, we observed negligible hole transfer except for the GA dimer where $p\approx 0.3$.
For all dimers, we found frequency content in the THz domain and noticed that Fourier amplitudes reflect $p$ and $\mathcal{P}_f$.
For dimers made of different monomers we generally get higher frequencies than for dimers made of identical monomers.

We compared the RT-TDDFT results with those obtained by TB in Refs.
\cite{Simserides:2014,LKGS:2014,LKMTLGThChS:2016} and we found that they are in good agreement.

\section*{Appendix}

\begin{figure*}[h!]
\centering
\includegraphics[width=7.5cm]{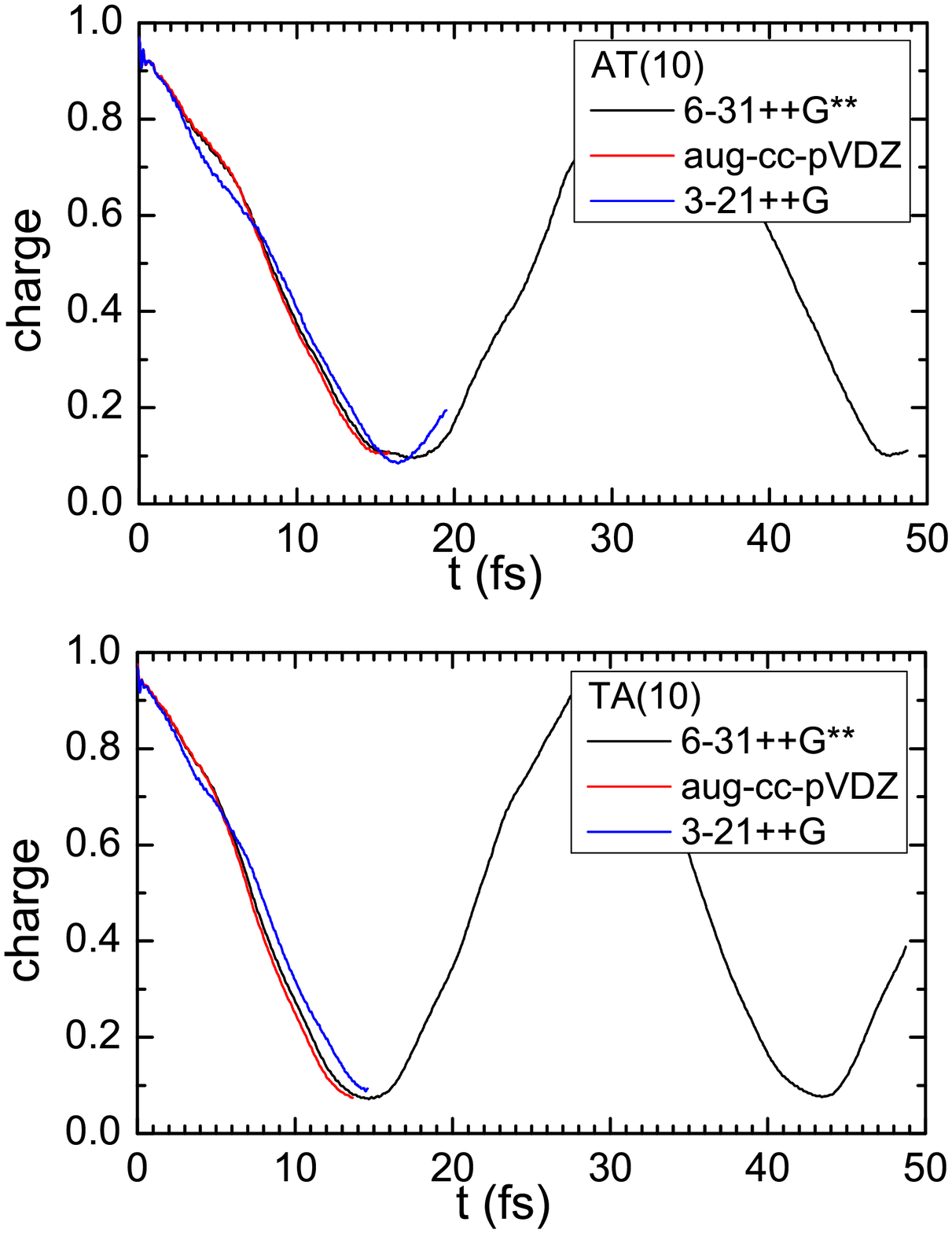}
\includegraphics[width=7.5cm]{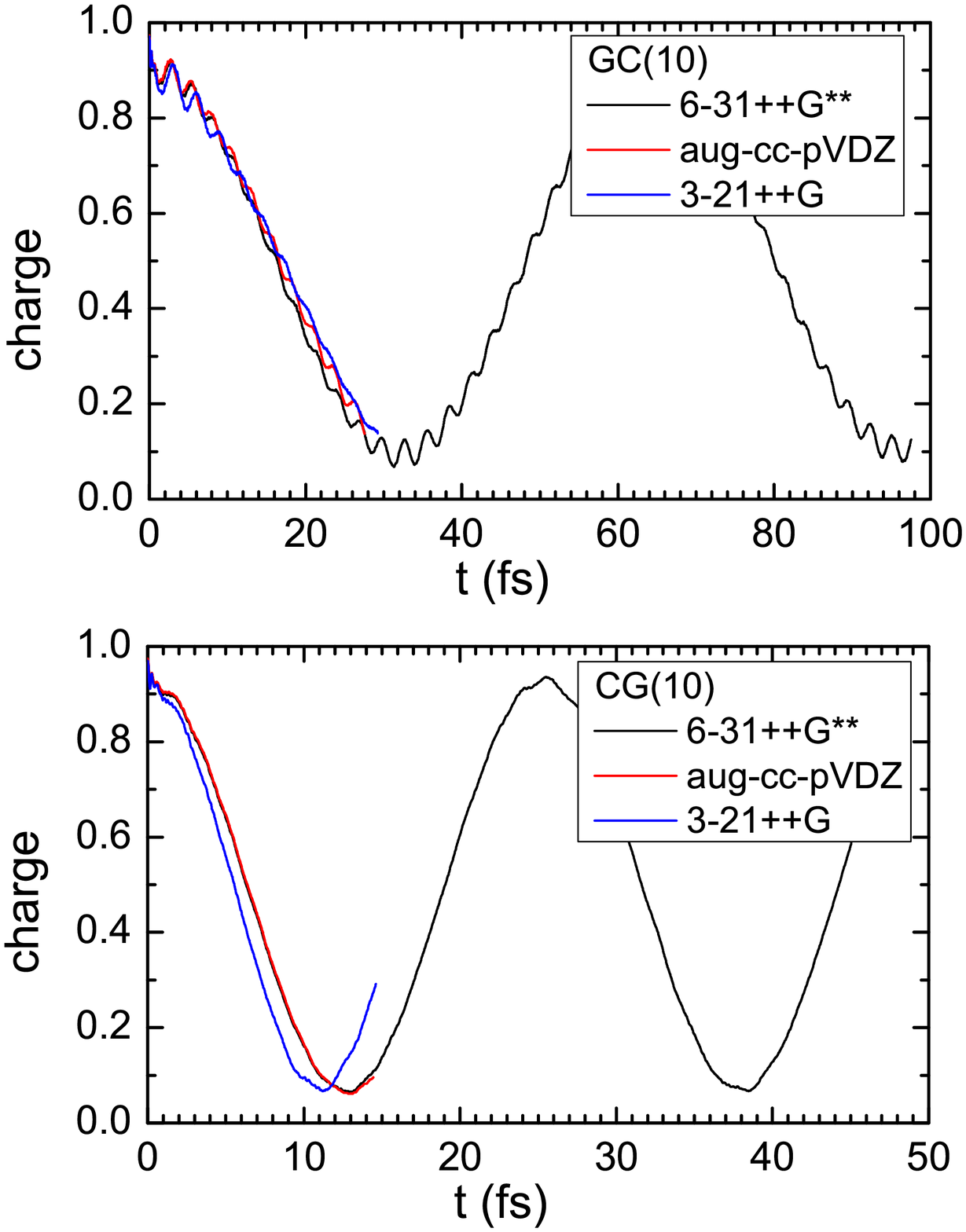}
\caption{Time evolution of hole transfer in dimers made of identical monomers, for the three different basis sets.}
\label{fig:same}
\end{figure*}

\begin{figure*}[h!]
\centering
\includegraphics[width=7.5cm]{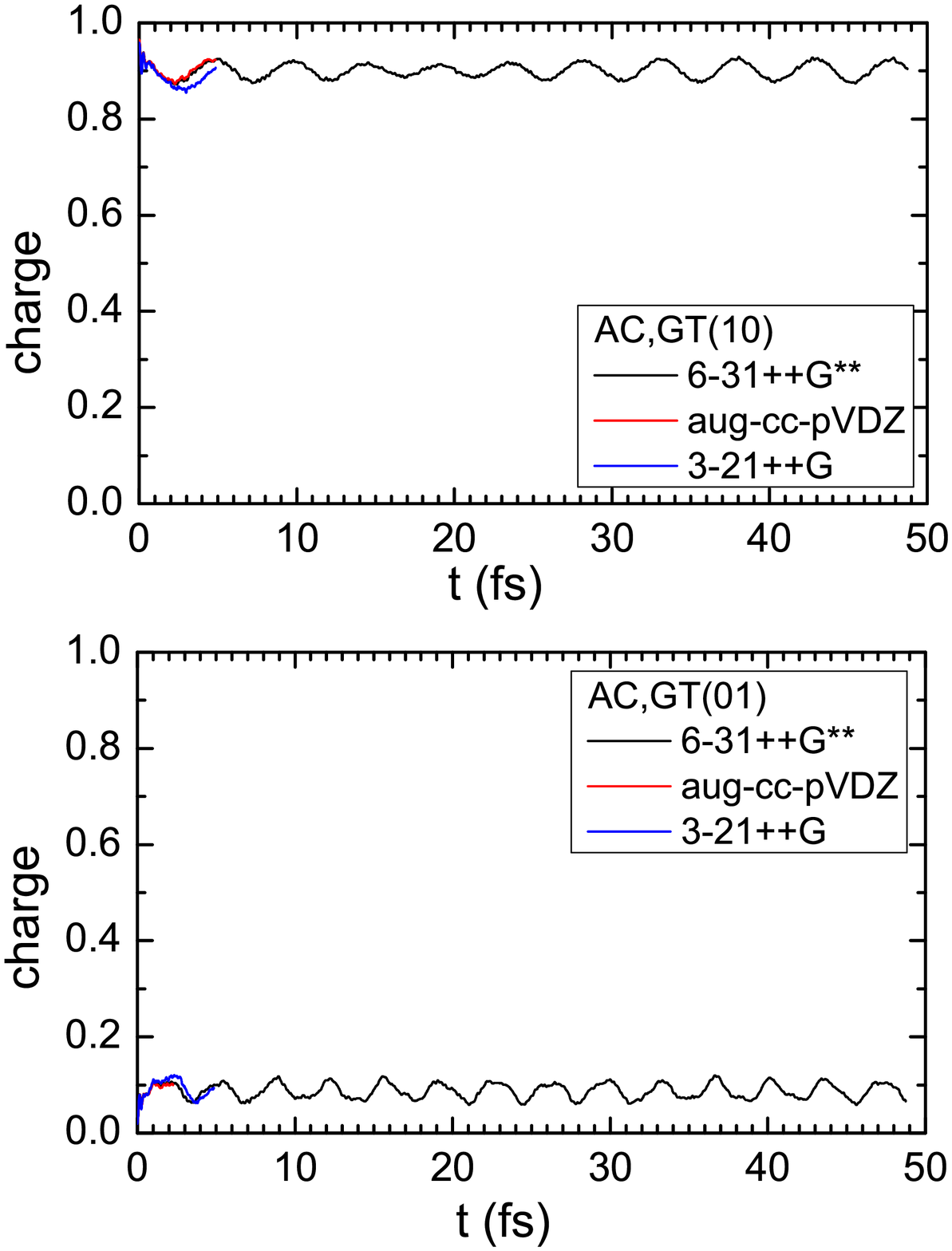}
\includegraphics[width=7.5cm]{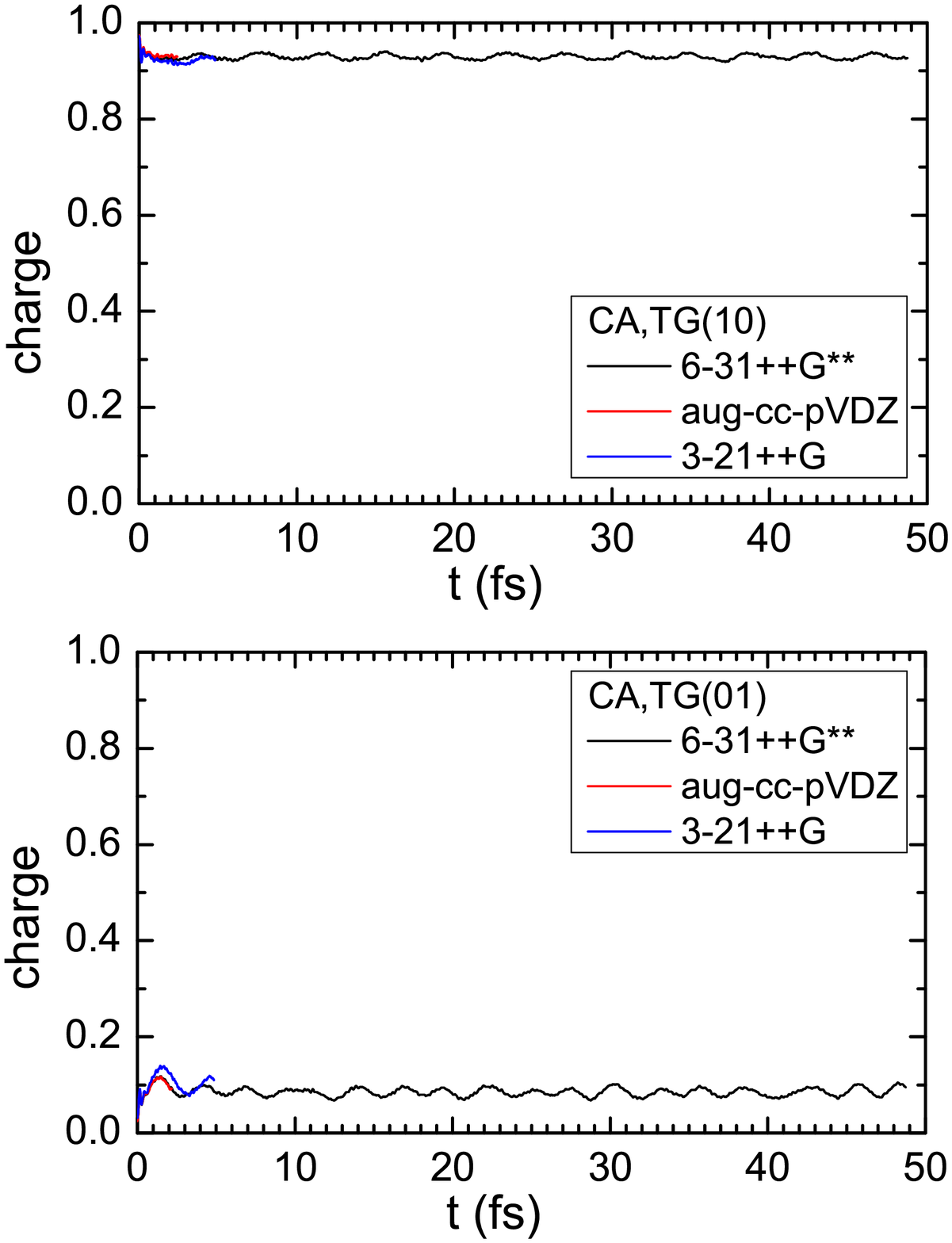}
\includegraphics[width=7.5cm]{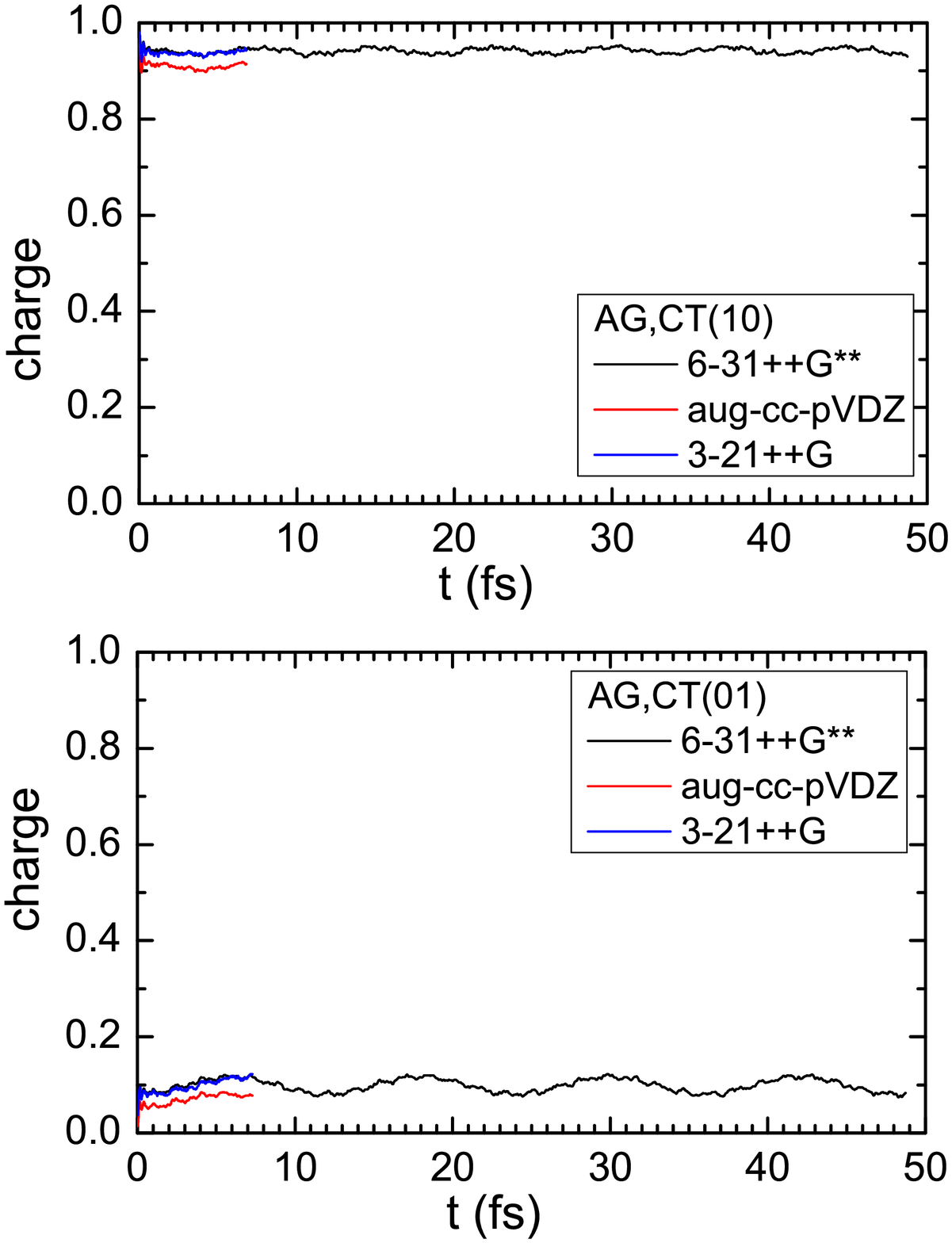}
\includegraphics[width=7.5cm]{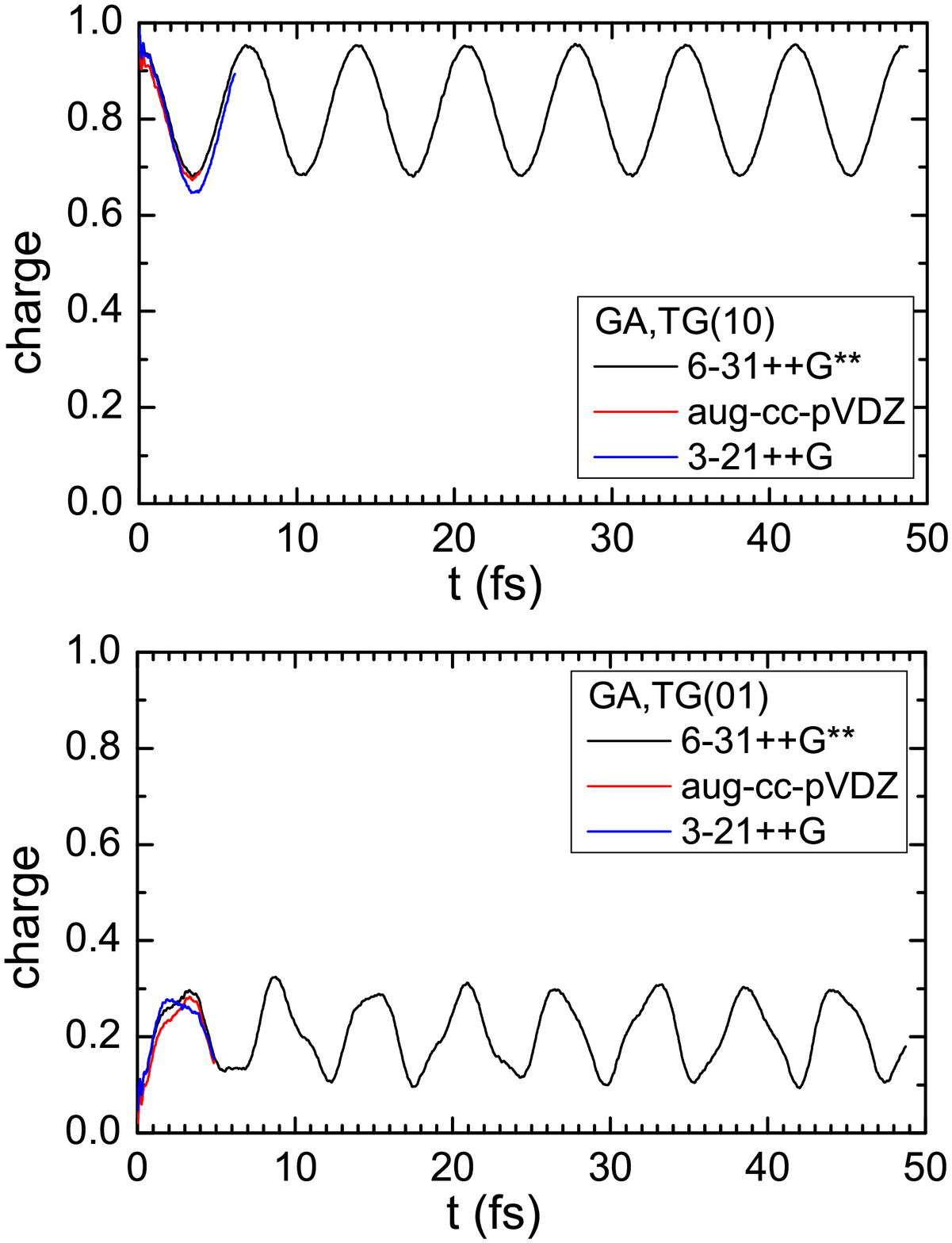}
\caption{Time evolution of hole transfer in dimers made of different monomers, for the three different basis sets.}
\label{fig:diff}
\end{figure*}

In Figs. \ref{fig:same}-\ref{fig:diff} we show charge oscillations calculated with different basis sets: 6-31++G**\cite{Hehre:1972}, 3-21++G \cite{Binkley:1980} and  aug-cc-pVDZ \cite{Dunning:1989} and we see that the results are similar (cf. Sec. \ref{sec:details}).
\clearpage

\clearpage

\section*{Acknowledgments}
This work was supported by computational time granted from the Greek Research \& Technology Network (GRNET) in the National HPC facility - ARIS - under project ID pr002008 - CODNA.

A. Morphis thanks the State Scholarships Foundation-IKY for a Ph.D. research scholarship via 'IKY Fellowships of Excellence', Hellenic Republic-Siemens Settlement Agreement.

\section*{Related work at}
\url{http://users.uoa.gr/~csimseri/physics_of_nanostructures_and_biomaterials.html}





\end{document}